# High performance integrated polarizers achieved by incorporating 2D layered graphene oxide films

Jiayang Wu[a], Yunyi Yang[a], Yuning Zhang[a], Yang Qu[a], Xingyuan Xu[a], Linnan Jia[a], Yao Liang[a], Sai T. Chu[b], Brent E. Little[c], Roberto Morandotti[d], Baohua Jia[a], and David Moss[a], *

[a]Centre for Micro-Photonics, Swinburne University of Technology, Hawthorn, VIC 3122, Australia
[b]City University of Hong Kong, Tat Chee Avenue, Hong Kong, China
[c]Xi'an Institute of Optics and Precision Mechanics Precision Mechanics of CAS, Xi'an, China
[d]INRS –Énergie, Matériaux et Télécommunications, Varennes, Québec, Canada

## ABSTRACT

Polarizers and polarization selective resonant cavities (e.g., ring resonators, gratings), are key components for applications to photography, coherent optical detection, polarization-division-multiplexing, optical sensing and liquid crystal displays. We demonstrate waveguide polarizers and polarization discriminating micro-ring resonators (MRRs) by integrating them with 2D graphene oxide (GO) layered thin films. We achieve precise control of the thickness, placement, and size of the films integrated onto photonic devices with a solution based, layer-by-layer transfer-free coating method combined with photolithography and lift-off. This overcomes limitations of layer transfer methods for 2D materials and is a significant advance to manufacturing integrated photonic devices incorporated with 2D materials. We measure the waveguide polarizer for different film thicknesses and lengths versus wavelength, polarization, and power, measuring a high polarization dependent loss (PDL) of ~ 53.8 dB. For GO-coated MRRs, we achieve an extinction ratio difference for TE/TM polarizations of 8.3-dB. We also present measurements of the linear optical properties of 2D layered GO films that yield the material loss anisotropy of the GO films and relative contribution of film loss anisotropy versus polarization-dependent mode overlap. Our results offer interesting physical insights into the transition of the layered GO films from 2D behaviour to quasi bulk like behavior and confirm the high performance of GO based integrated polarization selective devices.

## 1. INTRODUCTION

Polarization control is fundamental for many applications [1-5]. Integrated polarization-selective devices based on complementary metal-oxide-semiconductor (CMOS) compatible devices [1, 6, 7] offers advantages of compact footprint, mass producibility, scalability and stability [8-12]. Recently, the large optical anisotropy and wide wavelength response of 2D materials such as transition metal dichalcogenides and graphene have been recognized and used to realize polarization selective devices [13-16]. Generally, layer transfer processes [16-18] are used to integrate 2D materials onto CMOS compatible platforms. These require exfoliated or chemical vapour deposition grown 2D membranes that are then attached to dielectric substrates (silicon or silica wafers). Despite their wide use, transfer approaches are complex, making it difficult to realize accurate patterning or flexible placement, together with large-area continuous coating onto integrated wafers. Accurate control of the layer location, size and thickness is critical to optimize mode overlap and loss. Current methods limit the scale of incorporating 2D materials in integrated chips.

Graphene oxide (GO) is a highly promising 2D material [19-22] because of its easy preparation and tunability in its material properties. Recently [15], a broadband GO-polymer waveguide polarizer was reported with a PDL of ~ 40 dB, by introducing GO films onto polymer SU8 waveguides with drop-casting methods with a film thickness for each drop-casting step of ~ 0.5 μm and with a drop diameter of ~1.3 mm. These are not ideal to achieve accurate control of the location, length, and thickness of the films.

Recently [19], we reported transfer-free, large-area, high-quality coating of GO films onto waveguides with a solution-based method featuring layer-by-layer GO film deposition. In this paper, we demonstrate GO-coated polarization-selective micro-ring resonators (MRRs) and waveguide polarizers in a CMOS compatible platform [20]. We achieve precise control of the thickness, placement, and size of the GO films on photonic devices using the layer-by-layer method followed by lift-off and photolithography. The waveguide polarizer performance for different film thicknesses and lengths versus polarization, wavelength, and power, are measured, showing a very high polarization dependent loss (PDL) of ~ 53.8 dB. The coated MRRs show an 8.3-dB TE/TM polarization dependent loss. Our results verify the high-performance of integrated polarizers that incorporate layered 2D GO films.

## 2. WAVEGUIDE POLARIZERS

The waveguides were fabricated from Hydex glass surrounded by silica using CMOS compatible processes [23-25] with chemical mechanical polishing as the last step to remove the upper cladding to allow coating the top waveguide surface. A waveguide polarizer (Fig 1a) was uniformly coated with GO with a layer-by-layer solution-based method [19]. These 4 scalable steps for the depostion of GO monolayers were repeated to obtain multiple layer films.

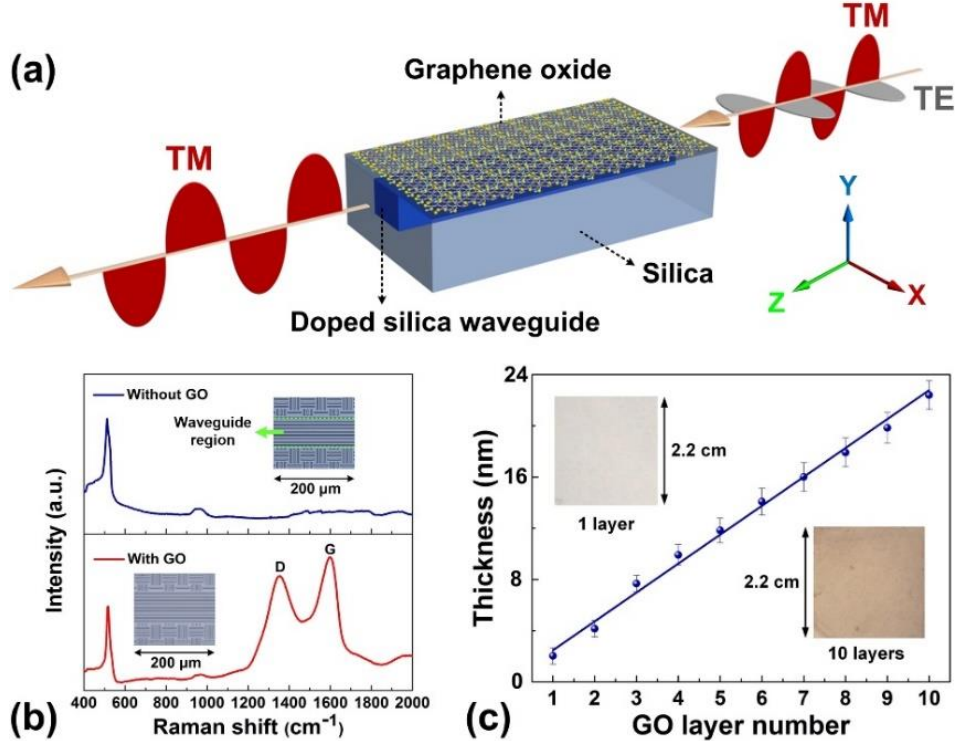

**Figure 1.** (a) Schematic illustration of GO-coated integrated waveguide polarizer. (b) Raman spectra of the integrated chip without GO and with 2 layers of GO. Insets show the corresponding microscope images with 9 parallel waveguides in the guiding region. (c) Measured GO film thickness versus GO layer number. Insets show the images of a 2.2 cm × 2.2 cm silica substrate coated with 1 and 10 layers of GO, respectively.

Uniform 1 - 10 layer coatings were deposited on waveguides. The Raman spectra of the waveguides (Figure 1(b)) without GO and with 2 layers of GO confirm the integration of GO onto the top surface with the observation of the G (1590 cm-1) and D (1345 cm-1) GO peaks. Microscope images of the waveguide with no layers and with 2 GO layers are (insets) illustrate the good film morphology. Figure 1(c) shows the film thickness vs number of layers measured with atomic force microscopy. The insets show images of 1 and 10 layers of GO coated on a 2.2 cm × 2.2 cm silica substrate with high uniformity. The film thickness vs layer number has a linear relationship showing an average thickness of ~2.18 nm per layer.

We also selectively patterned GO films with lithography and lift-off in order to achieve precise control of the dimensions and location of the GO films. This enabled us to fabricate GO-coated waveguide polarizers with a shorter length as well as thicker (up to 100) layers. The chip was spin-coated with photoresist and photolithographically patterned to open a window on the waveguide. Next, the films were deposited on the chip and patterned with this lift-off process.

An 8-channel single-mode fiber array was used to butt couple TE and TM polarized CW light from a tunable 1550 nm laser. The coupling loss between the fiber array and waveguides was about 8 dB/facet, which normally is ~1.0 dB/facet with the use of mode convertors [26, 27] that were absent here. The propagation loss of the uncoated 1.5-cm-long waveguides was very low (< 0.25 dB/cm) and so the total loss (TE= −16.2 dB; TM = −16.5 dB) of the uncoated devices was dominated by mode coupling loss. The slight PDL of the uncoated waveguides resulted from a slightly different mode-coupling mismatch as well as polarization dependent scattering from any residual roughness in the top polished surface.

We use two figures of merit (FOMs) – one for the excess propagation loss ($FOM_{EPL}$) and one for the overall polarization dependent loss ($FOM_{PDL}$) to characterize the performance of the devices:

$$FOM_{EPL} = (EPL_{TE} - EPL_{TM}) / EPL_{TM}, \qquad (1)$$
$$FOM_{PDL} = PDL / EIL, \qquad (2)$$

where the excess propagation losses, $EPL_{TE}$ (dB/cm) and $EPL_{TM}$ (dB/cm), are GO-induced excess propagation losses for the TE and TM polarizations, respectively. The polarization dependent loss (PDL (dB)) is defined as the ratio of the maximum to minimum insertion losses. The insertion loss induced by the GO film over the uncoated waveguide is defined as the excess insertion loss, EIL (dB), which considers only the insertion loss induced by GO and not the fibre-waveguide mode coupling loss or the propagation loss of the uncoated waveguide in the overall insertion loss. In our case, since the TM polarization had the lowest insertion loss, EIL is the excess GO-induced insertion loss for the TM polarization and is given by $EIL = EPL_{TM} \cdot L$, where L is the GO coating length. Note that $FOM_{EPL}$ only considers the difference in propagation loss due to the GO films, and so this is more accurate to characterize the material anisotropy, whereas $FOM_{PDL}$ is used more widely to evaluate the device performance [16] since it also includes background PDL of the uncoated waveguides. $FOM_{EPL}$ equals $FOM_{PDL}$ only when the TE and TM polarized insertion losses of the uncoated waveguide are the equal.

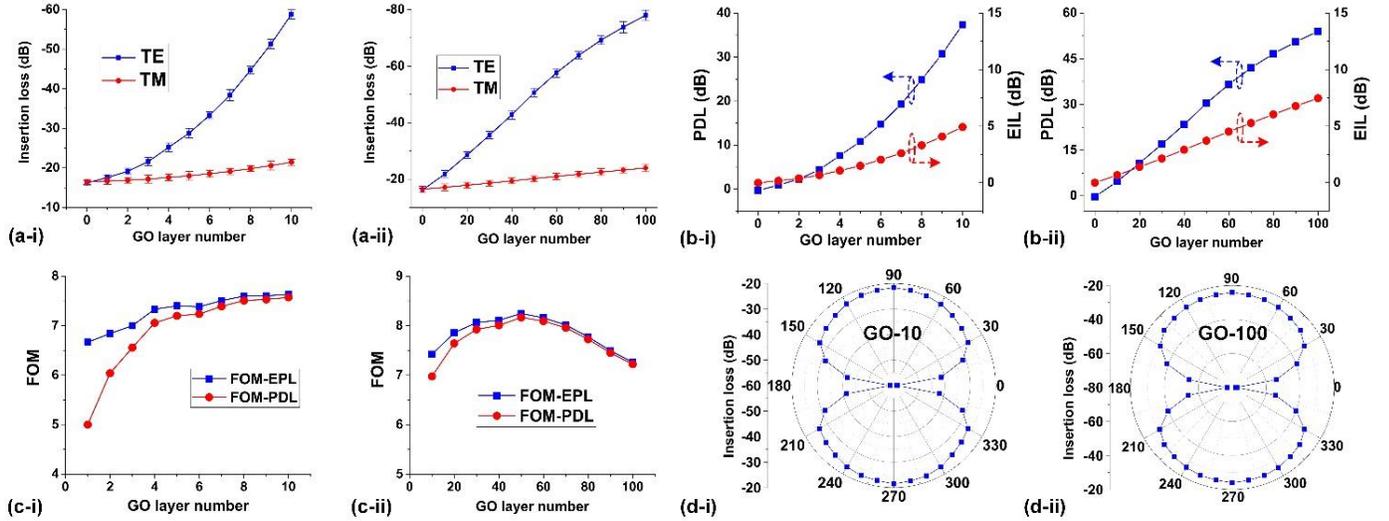

**Figure 2.** (a) Measured TE and TM polarized insertion loss. (b) Extracted polarization dependent loss (PDL) and excess insertion loss (EIL). (c) Calculated figure of merits (FOMs). In (a) − (c), (i) shows the results for 1.5-cm-long uniformly coated waveguides (0, 1, 2, ..., 10 layers of GO) and (ii) shows the results for 1.5-cm-long waveguides with 2-mm-long patterned GO (0, 10, 20, ..., 100 layers of GO). (d) Polar diagrams presenting the polarizer performance for (i) 1.5-cm GO coating length, 10 layers of GO and (ii) 2-mm GO coating length, 100 layers of GO. The polar angle represents the angle between the input polarization plane and the substrate. The input CW power and wavelength in (a) and (d) are 0 dBm and 1550 nm, respectively.

Figure 2 shows the polarization dependent performance (TE (in-plane) and TM (out-of-plane)) for both the 1.5-cm-long uniformly coated waveguides (0-10 layers), left side (i), and patterned (2-mm-long) devices (10-100 layers), right side (ii). Figure 2(a) shows the polarization dependent insertion loss, where the data depicts the average values of measurements for three identical devices. Figure 2(b) shws PDL and EIL calculated using the average insertion loss in Figure 2(a). Figures 2(c) and (d) show the FOMs calculated from Figure 2(b) and polar diagrams of the loss.

The TE insertion loss increases much more strongly with layer number than TM, thus showing a large PDL with low EIL and resulting in high-performance polarization selective devices. The maximum PDL was ~37.4 dB for 10 uniformly coated layers device and ~53.8 dB for a patterned device with 100 layers, with a moderate EIL of ~5.0 dB and ~7.5 dB for the two devices. By achieving better mode overlap with the films through optimizing the waveguide geometry, the EIL was reduced even more. Further, the PDL still increased at 2-3 dB/cm/layer at a thickness of 100 layers, and so much higher PDL can be achieved with thicker layers than 200 nm. Both $FOM_{EPL}$ and $FOM_{PDL}$ increase with a maximum of $FOM_{EPL}$ and $FOM_{PDL}$ of ~8.2 and ~8.1, respectively, at 50 layers, with their difference then decreasing. This is because the impact of the background PDL (~0.3 dB) became smaller as the EIL increased with increasing number of GO layers.

## 3. MICRO-RING RESONATORS

We coated integrated MRRs with GO films for polarization-selective MRRs for coherent receivers [28]. Figure 3(a) shows the coated MRRs, with the atomic GO structure (insets) and a scanning electron microscope image of the GO film with up to five GO layers. The unclad MRRs in Hydex glass were fabricated with the same CMOS compatible processes as the waveguides [29 - 40], with the ring and bus waveguides having the same dimensions. The MRR radius was ~592 µm, for a FSR of ~0.4 nm (~50 GHz), with a ring to bus waveguide gap of ~0.8 µm. Two GO-coated MRR polarizers were fabricated and tested - uniformly coated 1−5 layer GO coatings and 50 µm long patterned films with 10−100 layers. Gold markers were patterned with metal lift-off after photolithography and e-beam evaporation, were used for precise alignment and accurate film placement. Microscope images of the uniformly coated MRRs with 5 layers of GO and patterned with 50 layers of GO are shown in Figures 3(b) and (c). Although a number of concentric rings are shown, only the center ring was coupled with the through/drop bus waveguides to form a MRR − the rest were used for easy identification by eye.

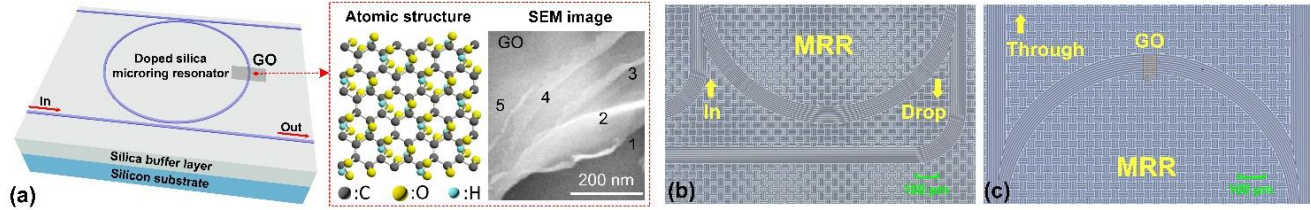

Figure 3. (a) Illustration of GO-based polarization-selective MRR. Insets show schematic atomic structure of GO and a scanning electron microscope (SEM) image of layered GO film. The numbers in the SEM refer to the number of layers for that part of the image. (b)−(c) Microscope images of the integrated MRR uniformly coated with 5 layers of GO and patterned with 50 layers of GO, respectively.

The uniformly GO-coated MRRs TE and TM polarized transmission spectra are shown in Figures 4(a-i) and (a-ii), respectively. The patterned MRR transmission spectra are shown in Figures 4(b-i) and (b-ii), measured with a CW power of 1mW. The Q factors [41-43] and extinction ratios [44, 45] are shown in Figure 5(a). The uncoated MRRs had high extinction ratios (> 15 dB) and relatively high Q's (180,000) (although less than buried waveguides) for both polarizations. This decreased with GO layer number – particularly for TE, as expected. For the patterned MRR with 50 GO layers, a maximum polarization extinction ratio (difference between the extinction ratios for the TE and TM polarized resonances) of ~8.3 dB was achieved. This can be improved by optimizing the GO film thickness, waveguide geometry, and coating size to balance material anisotropy and mode overlap. The hybrid integrated waveguide propagation loss for TE and TM polarizations was extracted with a scattering matrix method [46-49] to fit the measured spectra in Figures 4(a) and 4(b) and is shown for uniformly coated (0−5 layers) and patterned rings (0−100 layers) in Figures 5(b-i) and (b-ii), respectively, along with the measured waveguide propagation loss. Since the resonances did not vary over small wavelength ranges, we only fit one resonance around 1549.5 nm. The fit coupling coefficients between the ring and the bus waveguide for TE and TM polarizations were ~0.241 and ~0.230, respectively. The close agreement reflects the reproducibility and stability of the film coating method. The propagation loss of the ring resonators is slightly higher than the waveguides, particularly for TE. This is probably due to photo-thermal reduction of GO in the ring resonator at higher intensities [50].

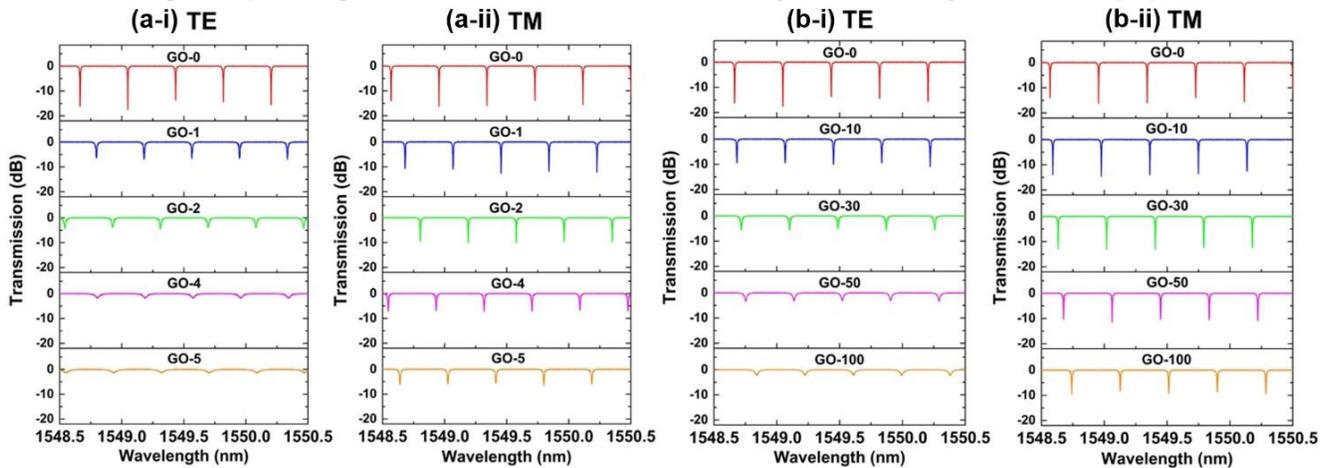

Figure 4. Measured transmission spectra of the MRR (a) uniformly coated with 0−5 layers and (d) patterned with 0−100 layers of GO.

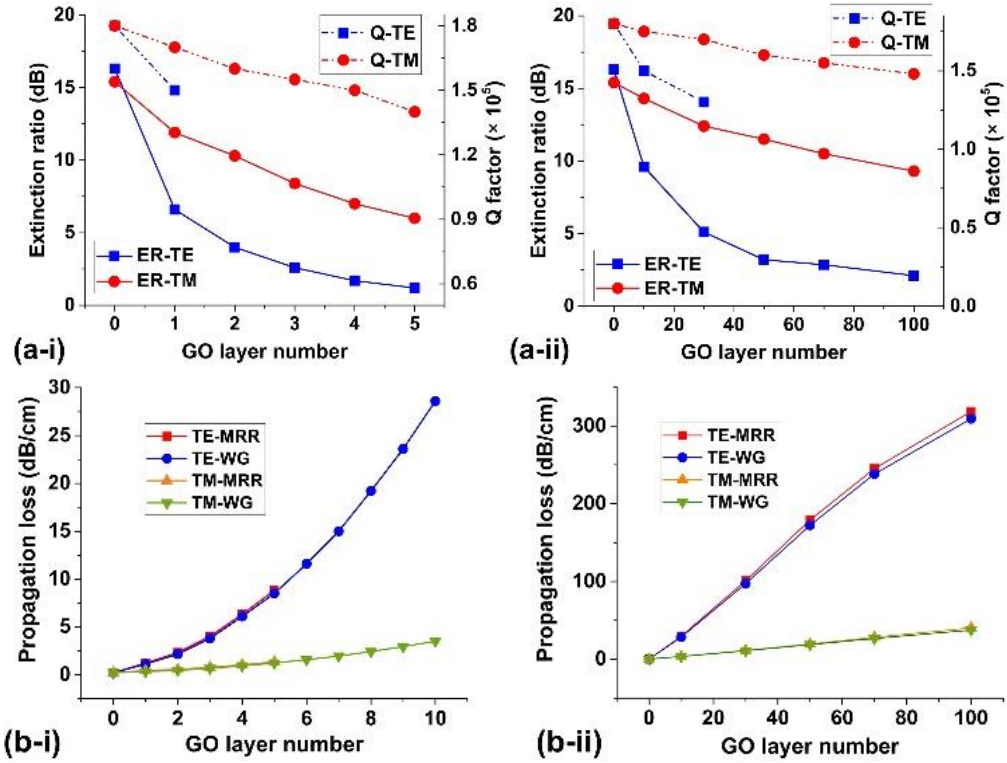

Figure 5. (a) Fit extinction ratios (ERs) and Q factors (Q) for the MRR (i) uniformly coated with 0−5 layers of GO and (ii) patterned with 0−100 layers of GO. The Q factors are not shown when the ER is < 5 dB. (b) Fit propagation loss obtained from the MRR experiment and waveguide propagation loss. (i) 0−10 layers of GO, (ii) 0−100 layers of GO.

## 4. CONCLUSION

We report polarization-selective MRRs and waveguide polarizers by integrating them with 2D GO films. We achieve precise control of the thickness, position, and size of the films via a layer-by-layer coating method along with photolithography and lift-off. We achieve a high PDL of ~53.8 dB for patterned waveguides and ~8.3-dB polarization extinction ratio between the TE and TM MRR resonances. The PDL is dominated by the material GO film loss anisotropy for thin films, and by polarization dependent mode overlap for thick films. These hybrid waveguide polarizers and polarization-selective MRRs are a powerful way to achieve high-performance polarization selective devices for PICs and complement recent work on nonlinear optics in GO integrated circuits [51-55].

## ACKNOWLEDGEMENTS

This work was supported by the Australian Research Council Discovery Projects Programs (No. DP150102972 and DP150104327) and the Swinburne ECR-SUPRA program. We also acknowledge the Swinburne Nano Lab for the support in device fabrication and characterization. RM acknowledges support by the Natural Sciences and Engineering Research Council of Canada (NSERC) through the Strategic, Discovery and Acceleration Grants Schemes, by the MESI PSR-SIIRI Initiative in Quebec, and by the Canada Research Chair Program.

## REFERENCES

[1] D. X. Dai, J. Bauters, and J. E. Bowers, "Passive technologies for future large-scale photonic integrated circuits on silicon: polarization handling, light non-reciprocity and loss reduction," *Light-Science & Applications,* vol. 1, Mar, 2012.
[2] D. X. Dai, L. Liu, S. M. Gao, D. X. Xu, and S. L. He, "Polarization management for silicon photonic integrated circuits," *Laser & Photonics Reviews,* vol. 7, no. 3, pp. 303-328, May, 2013.

[3] X. Xu, J. Wu, L. Jia, M. Tan, T. G. Nguyen, S. T. Chu, B. E. Little, R. Morandotti, A. Mitchell, and D. J. Moss, "Continuously tunable orthogonally polarized RF optical single sideband generator based on micro-ring resonators," *Journal of Optics,* vol. 20, no. 11, pp. 115701, 2018.

[4] X. Xu, J. Wu, M. Tan, T. G. Nguyen, S. Chu, B. Little, R. Morandotti, A. Mitchell, and D. J. Moss, "Orthogonally polarized RF optical single sideband generation and dual-channel equalization based on an integrated micro-ring resonator," *Journal of Lightwave Technology*, Vol. 36, No. 20, 4808-4818, 2018.

[5] Y. Zhang, Y. He, J. Wu, X. Jiang, R. Liu, C. Qiu, X. Jiang, J. Yang, C. Tremblay, and Y. Su, "High-extinction-ratio silicon polarization beam splitter with tolerance to waveguide width and coupling length variations," *Opt Express,* vol. 24, no. 6, pp. 6586-93, Mar 21, 2016.

[6] L. Lu, J. Wu, T. Wang, and Y. Su, "Compact all-optical differential-equation solver based on silicon microring resonator," *Frontiers of Optoelectronics,* vol. 5, no. 1, pp. 99-106, 2012.

[7] L. Zhang, J. Wu, X. Yin, X. Sun, P. Cao, X. Jiang, and Y. Su, "A High-Speed Second-Order Photonic Differentiator Based on Two-Stage Silicon Self-Coupled Optical Waveguide," *IEEE Photonics Journal,* vol. 6, no. 2, pp. 1-5, 2014.

[8] J. Wu, P. Cao, X. Hu, X. Jiang, T. Pan, Y. Yang, C. Qiu, C. Tremblay, and Y. Su, "Compact tunable silicon photonic differential-equation solver for general linear time-invariant systems," *Opt Express,* vol. 22, no. 21, pp. 26254-64, Oct 20, 2014.

[9] J. Wu, P. Cao, T. Pan, Y. Yang, C. Qiu, C. Tremblay, and Y. Su, "Compact on-chip 1 × 2 wavelength selective switch based on silicon microring resonator with nested pairs of subrings," *Photonics Research,* vol. 3, no. 1, pp. 9, 2014.

[10] S. Lai, Z. Xu, B. Liu, and J. Wu, "Compact silicon photonic interleaver based on a self-coupled optical waveguide," *Appl Opt,* vol. 55, no. 27, pp. 7550-5, Sep 20, 2016.

[11] J. Wu, B. Liu, J. Peng, J. Mao, X. Jiang, C. Qiu, C. Tremblay, and Y. Su, "On-Chip Tunable Second-Order Differential-Equation Solver Based on a Silicon Photonic Mode-Split Microresonator," *Journal of Lightwave Technology,* vol. 33, no. 17, pp. 3542-3549, 2015.

[12] L. Lu, F. Li, M. Xu, T. Wang, J. Wu, L. Zhou, and Y. Su, "Mode-Selective Hybrid Plasmonic Bragg Grating Reflector," *IEEE Photonics Technology Letters,* vol. 24, no. 19, pp. 1765-1767, 2012.

[13] Q. Bao, H. Zhang, B. Wang, Z. Ni, C. H. Y. X. Lim, Y. Wang, D. Y. Tang, and K. P. Loh, "Broadband graphene polarizer," *Nature Photonics,* vol. 5, no. 7, pp. 411-415, 2011.

[14] J. T. Kim, and C. G. Choi, "Graphene-based polymer waveguide polarizer," *Optics Express,* vol. 20, no. 4, pp. 3556-3562, Feb 13, 2012.

[15] W. H. Lim, Y. K. Yap, W. Y. Chong, C. H. Pua, N. M. Huang, R. M. De La Rue, and H. Ahmad, "Graphene oxide-based waveguide polariser: from thin film to quasi-bulk," *Opt Express,* vol. 22, no. 9, pp. 11090-8, May 5, 2014.

[16] H. Lin, Y. Song, Y. Huang, D. Kita, S. Deckoff-Jones, K. Wang, L. Li, J. Li, H. Zheng, Z. Luo, H. Wang, S. Novak, A. Yadav, C.-C. Huang, R.-J. Shiue, D. Englund, T. Gu, D. Hewak, K. Richardson, J. Kong, and J. Hu, "Chalcogenide glass-on-graphene photonics," *Nature Photonics,* vol. 11, no. 12, pp. 798-805, 2017.

[17] Y. Yang, R. Liu, J. Wu, X. Jiang, P. Cao, X. Hu, T. Pan, C. Qiu, J. Yang, Y. Song, D. Wu, and Y. Su, "Bottom-up Fabrication of Graphene on Silicon/Silica Substrate via a Facile Soft-hard Template Approach," *Sci Rep,* vol. 5, pp. 13480, Aug 27, 2015.

[18] X. Sun, C. Qiu, J. Wu, H. Zhou, T. Pan, J. Mao, X. Yin, R. Liu, W. Gao, Z. Fang, and Y. Su, "Broadband photodetection in a microfiber-graphene device," *Opt Express,* vol. 23, no. 19, pp. 25209-16, Sep 21, 2015.

[19] Y. Yang, J. Wu, X. Xu, Y. Liang, S. T. Chu, B. E. Little, R. Morandotti, B. Jia, and D. J. Moss, "Enhanced four-wave mixing in waveguides integrated with graphene oxide," *APL Photonics,* vol. 3, no. 12, pp. 120803, 2018.

[20] J. Wu, Y.Yang, Y. Qu, X. Xu, Y. Liang, S. T. Chu, B. E. Little, R. Morandotti, B. H. Jia, and D. J. Moss, "Graphene Oxide Waveguide and Micro-Ring Resonator Polarizers," *Laser & Photonics Reviews,* vol. 13, no. 9, 1900056, Sep, 2019.

[21] X. Li, H. Ren, X. Chen, J. Liu, Q. Li, C. Li, G. Xue, J. Jia, L. Cao, A. Sahu, B. Hu, Y. Wang, G. Jin, and M. Gu, "Athermally photoreduced graphene oxides for three-dimensional holographic images," *Nat Commun,* vol. 6, pp. 6984, Apr 22, 2015.

[22] X. Zheng, B. Jia, H. Lin, L. Qiu, D. Li, and M. Gu, "Highly efficient and ultra-broadband graphene oxide ultrathin lenses with three-dimensional subwavelength focusing," *Nat Commun,* vol. 6, pp. 8433, Sep 22, 2015.

[23] X. Xu, J. Wu, M. Shoeiby, T. G. Nguyen, S. T. Chu, B. E. Little, R. Morandotti, A. Mitchell, and D. J. Moss, "Reconfigurable broadband microwave photonic intensity differentiator based on an integrated optical frequency comb source," *APL Photonics,* vol. 2, no. 9, pp. 096104, 2017.

[24] X. Xu, J. Wu, T. G. Nguyen, S. T. Chu, B. E. Little, R. Morandotti, A. Mitchell, and D. J. Moss, "Broadband RF Channelizer based on an Integrated Optical Frequency Kerr Comb Source," *Journal of Lightwave Technology,* vol. 36, no. 19, pp. 4519-4526, 2018.

[25] J. Wu, X. Xu, T. G. Nguyen, S. T. Chu, B. E. Little, R. Morandotti, A. Mitchell, and D. J. Moss, "RF Photonics: An Optical Microcombs' Perspective," *IEEE Journal of Selected Topics in Quantum Electronics,* vol. 24, no. 4, pp. 1-20, 2018.

[26] X. Xu, J. Wu, T. G. Nguyen, T. Moein, S. T. Chu, B. E. Little, R. Morandotti, A. Mitchell, and D. J. Moss, "Photonic microwave true time delays for phased array antennas using a 49 GHz FSR integrated optical micro-comb source [Invited]," *Photonics Research,* vol. 6, no. 5, pp. B30, 2018.

[27] X. Xu, J. Wu, T. G. Nguyen, M. Shoeiby, S. T. Chu, B. E. Little, R. Morandotti, A. Mitchell, and D. J. Moss, "Advanced RF and microwave functions based on an integrated optical frequency comb source," *Opt Express,* vol. 26, no. 3, pp. 2569-2583, Feb 5, 2018.

[28] Y. Tan, S. Chen, and D. Dai, "Polarization-selective microring resonators," *Opt Express,* vol. 25, no. 4, pp. 4106-4119, Feb 20, 2017.
[29] X. Xu, M. Tan, J. Wu, T. G. Nguyen, S. Chu, B. Little, R. Morandotti, A. Mitchell, and D. J. Moss, "Advanced adaptive photonic RF filters with 80 taps based on an integrated optical micro-comb source," *Journal of Lightwave Technology* vol. 37, no. 4, 1288-1295, 2019.
[30] X. Xu, M. Tan, J. Wu, T. G. Nguyen, S. T. Chu, B. E. Little, R. Morandotti, A. Mitchell, and D. J. Moss, "High performance RF filters via bandwidth scaling with Kerr micro-combs," *APL Photonics,* vol. 4, no. 2, pp. 026102, 2019.
[31] A. Pasquazi, M. Peccianti, Y. Park, B. E. Little, S. T. Chu, R. Morandotti, J. Azaña, and D. J. Moss, "Sub-picosecond phase-sensitive optical pulse characterization on a chip", *Nature Photonics*, vol. 5, no. 10, pp. 618 - 623 (2011). DOI: 10.1038/nphoton.2011.199.
[32] H. Bao et al., "Laser cavity-soliton microcombs," *Nature Photonics,* vol. 13, no. 6, pp. 384-389 (2019).
[33] M. Kues, et. al., "Passively modelocked laser with an ultra-narrow spectral width", *Nature Photonics*, vol. 11, no. 3, p159 (2017). DOI:10.1038/nphoton.2016.271
[34] A.Pasquazi et al., "Micro-Combs: A Novel Generation of Optical Sources", *Physics Reports*, vol. 729, no. 1, 1-81 (2018).
[35] L. Razzari, D. Duchesne, M. Ferrera et al., "CMOS-compatible integrated optical hyper-parametric oscillator," *Nature Photonics,* vol. 4, no. 1, pp. 41-45 (2010).
[36] M. Ferrera, L. Razzari, D. Duchesne et al., "Low-power continuous-wave nonlinear optics in doped silica glass integrated waveguide structures," *Nature Photonics*, vol. 2, no. 12, pp.737-740 (2008).
[37] C. Reimer et al., "High-dimensional one-way quantum processing implemented on d-level cluster states", *Nature Physics, vol.* 15, no. 2, pp. 148–153 (2019). DOI:10.1038/s41567-018-0347-x
[38] C. Reimer, M. Kues, P. Roztocki et al., "Generation of multiphoton entangled quantum states by means of integrated frequency combs," *Science*, vol. 351, no. 6278, pp.1176-1180 (2016).
[39] M.Kues, et al., "On-chip generation of high-dimensional entangled quantum states and their coherent control", *Nature*, vol. 546, no. 7660, pp.622-626 (2017).
[40] M. Kues, C. Reimer, A. Weiner, J. Lukens, W. Munro, D. J. Moss, and R. Morandotti, "Quantum Optical Micro-combs", *Nature Photonics*, vol. 13, no. 3, pp. 170-179 (2019). DOI:10.1038/s41566-019-0363-0.
[41] J. Wu, T. Moein, X. Xu, G. Ren, A. Mitchell, and D. J. Moss, "Micro-ring resonator quality factor enhancement via an integrated Fabry-Perot cavity," *APL Photonics,* vol. 2, no. 5, pp. 056103, 2017.
[42] J. Wu, T. Moein, X. Xu, and D. J. Moss, "Advanced photonic filters based on cascaded Sagnac loop reflector resonators in silicon-on-insulator nanowires," *APL Photonics,* vol. 3, no. 4, pp. 046102, 2018.
[43] T. Pan, C. Qiu, J. Wu, X. Jiang, B. Liu, Y. Yang, H. Zhou, R. Soref, and Y. Su, "Analysis of an electro-optic modulator based on a graphene-silicon hybrid 1D photonic crystal nanobeam cavity," *Opt Express,* vol. 23, no. 18, pp. 23357-64, Sep 7, 2015.
[44] T. Wang, M. Xu, F. Li, J. Y. Wu, L. J. Zhou, and Y. K. Su, "Design of a high-modulation-depth, low-energy silicon modulator based on coupling tuning in a resonance-split microring," *Journal of the Optical Society of America B-Optical Physics,* vol. 29, no. 11, pp. 3047-3056, Nov, 2012.
[45] X. Mu, W. Jiayang, W. Tao, H. Xiaofeng, J. Xinhong, and S. Yikai, "Push–Pull Optical Nonreciprocal Transmission in Cascaded Silicon Microring Resonators," *IEEE Photonics Journal,* vol. 5, no. 1, pp. 2200307-2200307, 2013.
[46] X. Jiang, J. Wu, Y. Yang, T. Pan, J. Mao, B. Liu, R. Liu, Y. Zhang, C. Qiu, C. Tremblay, and Y. Su, "Wavelength and bandwidth-tunable silicon comb filter based on Sagnac loop mirrors with Mach-Zehnder interferometer couplers," *Opt Express,* vol. 24, no. 3, pp. 2183-8, Feb 8, 2016.
[47] J. Wu, P. Cao, X. Hu, T. Wang, M. Xu, X. Jiang, F. Li, L. Zhou, and Y. Su, "Nested Configuration of Silicon Microring Resonator With Multiple Coupling Regimes," *IEEE Photonics Technology Letters,* vol. 25, no. 6, pp. 580-583, 2013.
[48] J. Wu, X. Jiang, T. Pan, P. Cao, L. Zhang, X. Hu, and Y. Su, "Non-blocking 2 × 2 switching unit based on nested silicon microring resonators with high extinction ratios and low crosstalks," *Chinese Science Bulletin,* vol. 59, no. 22, pp. 2702-2708, 2014.
[49] J. Wu, J. Peng, B. Liu, T. Pan, H. Zhou, J. Mao, Y. Yang, C. Qiu, and Y. Su, "Passive silicon photonic devices for microwave photonic signal processing," *Optics Communications,* vol. 373, pp. 44-52, 2016.
[50] W. Y. Chong, W. H. Lim, Y. K. Yap, C. K. Lai, R. M. De La Rue, and H. Ahmad, "Photo-induced reduction of graphene oxide coating on optical waveguide and consequent optical intermodulation," *Sci Rep,* vol. 6, pp. 23813, Apr 1, 2016.
[51] Y. Zhang et al., "Enhanced Kerr nonlinearity and nonlinear figure of merit in silicon nanowires integrated with 2D graphene oxide films", *ACS Applied Materials and Interfaces*, vol. 12, no. 29, pp. 33094−33103 (2020). DOI:10.1021/acsami.0c07852
[52] Y. Qu et al., "Enhanced nonlinear four-wave mixing in silicon nitride waveguides integrated with 2D layered graphene oxide films", *Advanced Optical Materials*, vol. 8, no. 20, 2001048 (2020). DOI: 10.1002/adom.202001048
[53] J. Wu et al., "Enhanced nonlinear four-wave mixing in microring resonators integrated with layered graphene oxide films", *Small,* vol. 16, no. 16, 1906563 (2020). DOI: 10.1002/smll.201906563
[54] D. Moss, "Graphene oxide films for advanced ultra-flat optical devices and photonic integrated circuits", TechRxiv (2020). DOI:10.36227/techrxiv.12980222.v1
[55] Wu, J.; Jia, L.; Zhang, Y.; Qu, Y.; Jia, B.; Moss, D.J., "Versatile Graphene Oxide Films for Ultra-flat Optics and Integrated Photonic Chips", Preprints 2020, 2020090492.

Wu, J.; Jia, L.; Zhang, Y.; Qu, Y.; Jia, B.; Moss, D.J. Versatile Graphene Oxide Films for Ultra-flat Optics and Integrated Photonic Chips. Preprints 2020, 2020090492